\documentclass[prl,showpacs,preprintnumbers,twocolumn,showkeys,floatfix]{revtex4}
\usepackage{graphicx}

\renewcommand{\vec}[1]{\mbox{\boldmath$#1$}}
\def\ga{\mathrel{\mathchoice {\vcenter{\offinterlineskip\halign{\hfil
$\displaystyle##$\hfil\cr>\cr\sim\cr}}}
{\vcenter{\offinterlineskip\halign{\hfil$\textstyle##$\hfil\cr
>\cr\sim\cr}}}
{\vcenter{\offinterlineskip\halign{\hfil$\scriptstyle##$\hfil\cr
>\cr\sim\cr}}}
{\vcenter{\offinterlineskip\halign{\hfil$\scriptscriptstyle##$\hfil\cr
>\cr\sim\cr}}}}}
\def\la{\mathrel{\mathchoice {\vcenter{\offinterlineskip\halign{\hfil
$\displaystyle##$\hfil\cr<\cr\sim\cr}}}
{\vcenter{\offinterlineskip\halign{\hfil$\textstyle##$\hfil\cr  
<\cr\sim\cr}}}
{\vcenter{\offinterlineskip\halign{\hfil$\scriptstyle##$\hfil\cr
<\cr\sim\cr}}}
{\vcenter{\offinterlineskip\halign{\hfil$\scriptscriptstyle##$\hfil\cr
<\cr\sim\cr}}}}}
\begin{document}
\preprint{APS/123-QED}
\title{Role of soft-iron impellers on the mode selection in
  the VKS dynamo experiment}
\author{Andr\'e Giesecke}
\email{a.giesecke@fzd.de}
\author{Frank Stefani}%
\author{Gunter Gerbeth}
\affiliation{
Forschungszentrum Dresden Rossendorf
}
\date{\today}
\begin{abstract}
A crucial point for the understanding of the 
von-K\'arm\'an-Sodium (VKS) dynamo experiment 
is the influence of soft-iron impellers. 
We present numerical simulations of a VKS-like dynamo 
with a localized permeability distribution that
resembles the shape of the flow driving impellers.
It is shown that the presence of soft-iron
material essentially determines the dynamo 
process in the VKS experiment.
An axisymmetric magnetic field mode can be explained by the combined action of
the soft-iron disk and a rather small $\alpha$-effect parametrizing the
induction effects of unresolved small scale flow fluctuations.  
\end{abstract}
\maketitle
Homogenous dynamo action, i.e. the transformation
of the kinetic energy of a flowing conducting fluid
into magnetic energy, accounts for the generation of 
planetary, stellar and galactic magnetic fields.
The last decade has seen tremendous progress in the experimental investigation of 
dynamo action \cite{ISI:000262272000001}.
By exploring parameter regions that
are hardly accessible to numerical simulations, such
experiments are essential for a  better 
understanding of geo- and astrophysical magnetic fields.
So far the threshold of magnetic field 
self-excitation has been crossed only in
three liquid sodium experiments  
in Riga, Karlsruhe, and Cadarache.
In the latter one, the von-K\'arm\'an-Sodium (VKS) 
experiment, a turbulent flow of liquid sodium is driven
by two counter-rotating impellers located at the 
opposing end caps of a
cylindrical domain, and field generation sets in when
the magnetic Reynolds number exceeds the critical value of
${\rm{Rm}}^{\rm{c}}\approx 32$ \cite{2007PhRvL..98d4502M}. 
Besides of the rich dynamical behavior, including 
periodic and irregular field reversals that occur when the impellers rotate
with different frequencies \cite{2007EL.....7759001B}, the structure of the
measured field attracts attention because of its high degree of axisymmetry
\cite{2009PhFl...21c5108M}.  
This geometry is in contrast to all numerical predictions based on
axisymmetric flows from which a magnetic field with an azimuthal wavenumber
$m=1$ ($B \propto\cos\varphi$) was expected \cite{2003EPJB...33..469M,
  2005PhFl...17k7104R, 2005physics..11149S}.   
Furthermore, dynamo action is
only obtained when the impellers are
made of soft-iron with a relative permeability 
of the order of $\mu_{\rm{r}}\approx 100$.
So far, there is neither a satisfying explanation of the 
observed axisymmetric field, nor of  
the failure of the dynamo in case that 
non-magnetic impellers are being used. This means, in turn, that 
the very working principle of the 
VKS dynamo is not understood at all.
A first attempt to explain the axisymmetric 
mode in the VKS experiment 
was recently made in terms of a turbulent $\alpha\omega$-model
\cite{2007GApFD.101..289P, 2008PhRvL.101j4501L}. 
The consideration of turbulence seems imperative
since the kinetic Reynolds number is of the order 
${\rm{Re}}\sim$$10^6$, and
flow and field exhibit strong fluctuations that are of the same
order as the mean values.
The model allows to circumvent the 
restrictions
imposed by Cowling's theorem which forbids 
axisymmetric
dynamo action
\cite{1933MNRAS..94...39C}.
Indeed, it was shown in \cite{andregafd} that an
axisymmetric VKS-like flow supported by an $\alpha$-effect is able to excite 
an axisymmetric field. 
However, the
magnitude of the $\alpha$-effect which is necessary 
to explain the dominant axial dipole turned out to 
be several times larger than the rms value of
the turbulent velocity.
Therefore, this simple $\alpha\omega$-model 
seems not capable of explaining the VKS dynamo mechanism.
But what about the role of the soft-iron impellers? 
Originally, their use was motivated
by the hope to shield off the field in the bulk of the cylinder 
from the lid layers behind the impellers.
These layers, and particularily any rotating flow therein, 
had been shown to be extremely detrimental for dynamo action 
\cite{2005physics..11149S}.   
First attempts to investigate the role of the high
permeability impellers 
utilized simplifying boundary conditions that enforce a
vanishing tangential field at the top and the bottom of 
the cylindrical
domain which mimics the field behaviour in case of 
end caps with infinite $\mu_{\rm{r}}$ \cite{2008EL.....8229001G, 2008PhRvL.101j4501L}.
A similar approach was followed in \cite{2009EL.....8739002G} with focus 
on the impact of an additional non-axisymmetric velocity component with
azimuthal wave number ($m=8$) that describes a hypothetic vortex structure between adjacent impeller blades. 
Comparable to the $\alpha\omega$-model with a localized $\alpha$-effect 
\cite{2008PhRvL.101j4501L, andregafd}, an axisymmetric field was obtained when
the amplitude of the distortion is of the same order as the mean flow.  
This means, however, that a similar problem as for the $\alpha\omega$-model
arises, because the assumed existence of highly correlated, large amplitude
non-axisymmetric motions could not be confirmed in water experiments
\cite{BURGUETEPERSONAL}. 
In this Letter, we will treat the rotating soft-iron
impellers in a more realistic manner
than just to assume
vanishing tangential field conditions at the end caps. 
We consider dynamo action 
caused by an axisymmetric mean flow
$\vec{u}$ which might be supported by an $\alpha$-effect.
The large scale magnetic flux density $\vec{B}$ is
governed by the mean field induction equation  
\begin{equation}
\frac{\partial\vec{B}}{\partial t}=
\nabla\times\left(\vec{u}\times\vec{B}+{{\alpha}}\vec{B}-\frac{1}{\mu_0\sigma(\vec{r})}
\nabla\times\frac{\vec{B}}{\mu_{\rm{r}}(\vec{r})}\right). 
\label{eq::general_induction}
\end{equation}
Here  $\sigma$ denotes the conductivity, $\mu_0$ the
permeability of vacuum and $\mu_{\rm{r}}$ the
relative permeability.
An $\alpha$-effect might be
caused either by a gradient of the turbulence
intensity 
or by helical fluid motions predominantly inbetween 
the eight blades of the impeller disks. 
In the latter case, the $(m=8)$ flow modulation is considered as a
small scale variation whose averaged influence is parametrized by an
axisymmetric $\alpha$-source term \cite{2008PhRvL.101j4501L, andregafd}.   
%}}
%
From Eq.~(\ref{eq::general_induction}) it follows 
that a high (even if only localized) permeability $\mu_{\rm{r}}$ reduces the effective magnetic
diffusivity $\eta^{\rm{eff}}=(\sigma\mu_0\mu_{\rm{r}}^{\rm{eff}})^{-1}$, where 
$\mu_{\rm{r}}^{\rm{eff}}=V^{-1}\int\mu_{\rm{r}}(\vec{r})d^3r$ and $V$ denotes the
cylinder volume.  
In addition, the terms on the 
r.h.s. of Eq.~(\ref{eq::general_induction})
that involve gradients of   
$\sigma$ or $\mu_{\rm{r}}$ are able
to couple toroidal and poloidal field
components in a way that possibly allows exponentially 
growing solutions for $\vec{B}$.
A paradigm for this type of 
dynamo has been presented by Busse and Wicht
\cite{1992gafd...64..135B} who showed that 
even a constant flow over an infinite plate with spatially varying
conductivity is able to sustain a dynamo. 
Interestingly, the structure of the field is then determined by a constant part
modulated by a contribution with the wave number of the conductivity variation. 
A similar effect results from the inhomogeneous 
permeability distribution in the VKS
experiment, and it is worthwhile to check if this
is already sufficient to explain 
a significant axisymmetric contribution. 
However, a purely axisymmetric field cannot be expected as such a solution
is forbidden even in case of space- and time dependent permeability/conductivity
distributions \cite{1982GApFD..19..301H}.

Eq.~(\ref{eq::general_induction}) is solved numerically applying a finite
volume method \cite{2008giesecke_maghyd}.
Note that at the experimental temperature of
$\sim 120^{\circ}\mbox{C}$, the differences of the electrical 
conductivity of iron
and liquid sodium are not very important so that
we focus on a non-uniform permeability distribution.
In the grid based approach the implementation of spatially varying
material coefficients is straightforward and only requires the application of
a simple averaging directive for the reconstruction of the local
(discretized) value of $\mu_{\rm{r}}$ that is seen by a field component 
at a certain grid cell.  
Although the scheme is easy to implement, the simulations are still
demanding because large gradients of $\mu_{\rm{r}}$ must be resolved. 
To keep the computational time in a reasonable limit, the presented results
are restricted to a maximum value of $\mu_{\rm{r}}=60$.
This is in rough agreement with recent measurements of the permeability of a
VKS-like impeller that gave values of  $\mu_{\rm{r}}\approx 65\dots 70$ \cite{gp_pc}.

\begin{figure}[h!]
\hspace*{-4.4cm}
\includegraphics[width=0.23\textwidth,height=5.4cm]{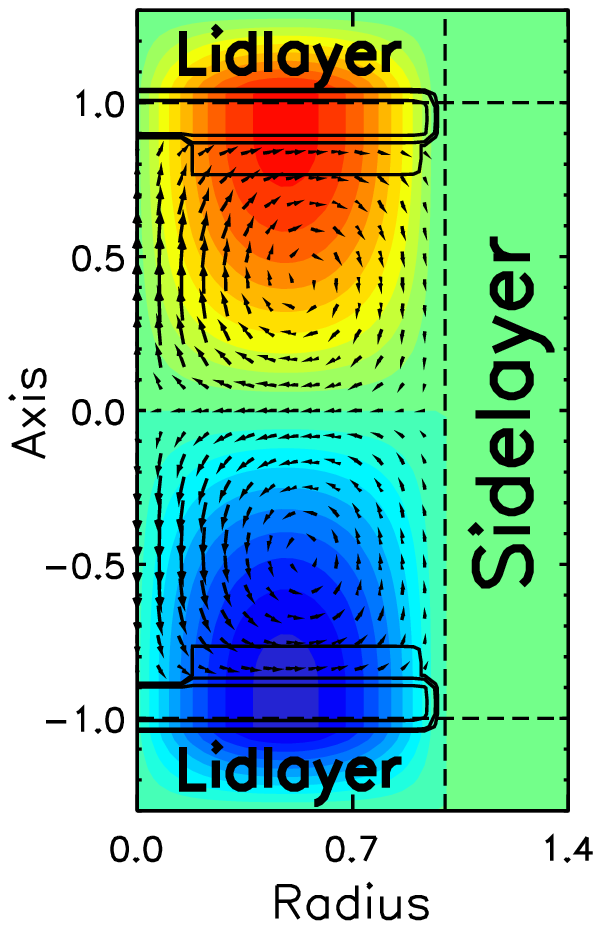}\nolinebreak[4!]
\\[-5.6cm]
\hspace*{4.5cm}
\includegraphics[width=0.23\textwidth,height=5.2cm]{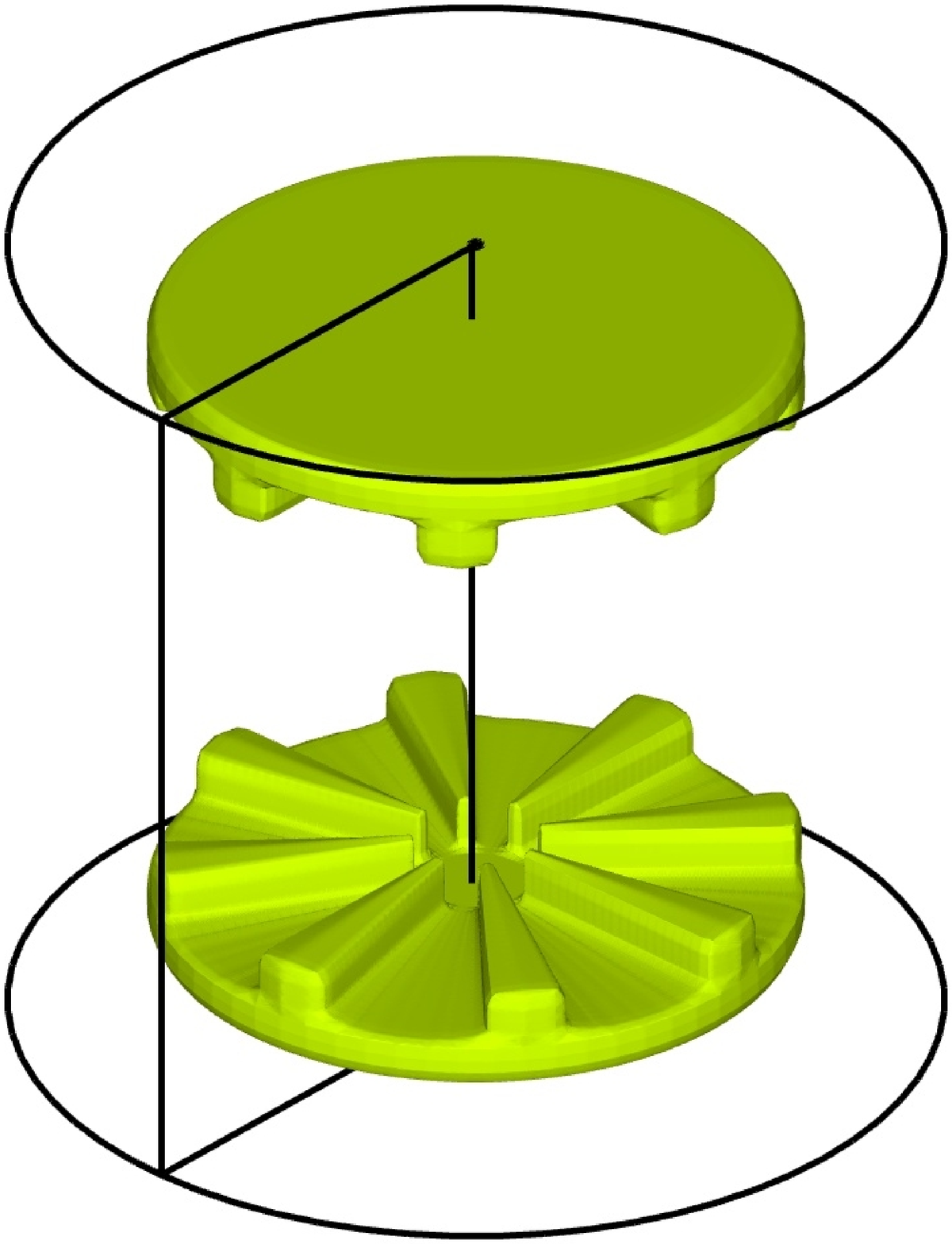}
\caption{Left panel: Structure of the prescribed axisymmetric velocity field. The color
  coded pattern represents the azimuthal velocity and the arrows show the
  poloidal velocity field. The black solid lines represent the shape of the
  impeller system (disk and blades). 
  Right panel: assumed permeability distribution; isolevel at $\mu_{\rm{r}}=35$ where the
  peak value is given by $\mu_{\rm{r}}=40$. In the fluid region $\mu_{\rm{r}}$ is
  equal to unity.}\label{fig::velfield}
\end{figure}

A suitable mean flow that describes the velocity distribution
in the VKS-experiment is represented by the so called MND-flow
\cite{2004phfl},
\begin{eqnarray}
u_r(r,z)&=&-\pi/H\cos\!\left({{2\pi z/H}}\right)r(1-r)^2(1+2r),\nonumber\\
u_{\varphi}(r,z)&=&4\epsilon r(1-r)\sin\left({{\pi z/H}}\right),\label{eq::s2t2}\\
u_z(r,z)&=&(1-r)(1+r-5r^2)\sin\left({{2\pi z/H}}\right)\nonumber,
\end{eqnarray}
where $H=1.8$ denotes the distance between both impeller disks and
$\epsilon$ describes the relation between toroidal and poloidal component of
the velocity. Here, $\epsilon=0.7259$ which is the optimum
value for obtaining dynamo action, at least 
when the mean flow alone is considered. 
The entire model setup is visualized in the left panel of
Fig.~\ref{fig::velfield}.  
The velocity field~(\ref{eq::s2t2}) is only applied in the region between the two
impellers.
The flow active region is restricted to a cylinder of radius $R=1$ (corresponding to
$20.5 \mbox{ cm}$ in the experiment) which is embedded in a layer of stagnant fluid with
a thickness of $0.4R$.
This so-called side layer significantly reduces ${\rm{Rm}}^{\rm{c}}$ \cite{2005physics..11149S}.
A lid layer is added behind each impeller disk.  
Within these lid layers a purely rotating flow is assumed, modeled by a linear
interpolation of the azimuthal velocity at the outer 
side of 
the impeller disk
towards to zero at the end cap of the cylindrical domain.
The presence of lid- and side layers reduces the influence of the magnetic
boundary conditions so that, for the sake of simplicity, vanishing tangential
field components are imposed at the outer surface of the computational domain.
The solid black lines in the left panel of Fig.~\ref{fig::velfield} represent
isolevels of the permeability distribution that 
mimics the impellers with radius $r=R$ and height $h=0.2$.
The impeller system is composed of two fractions: the disk, which
is modeled by an axisymmetric scalar field with constant
$\mu_{\rm{r}} \geq 1$, and an azimuthally varying contribution 
located adjacent to the inner
side of each disk.
This part represents the system of eight blades and is modeled as an 
azimuthal rotating scalar field $\propto \cos(8\varphi+\omega t)$ in a way that
the minimum value (located between the blades) corresponds to $\mu_{\rm{r}}=1$ (the
permeability of the fluid) and the peak value corresponds to the disk permeability.
Here, $\omega$ denotes the angular velocity of the impeller that is necessary to
drive a flow with a given ${\rm{Rm}}$ (defined as
${\rm{Rm}}=R\mu_0\sigma|U_{\max}|$).   
The 3d structure of the resulting permeability distribution is shown in
the right panel of Fig.~\ref{fig::velfield}. 
Note that the impellers are modeled only by the permeability distribution
and no particular flow boundary conditions are enforced on the (assumed)
interface between impeller and fluid.

Initially, simulations have been performed without any $\alpha$-effect.
After a transient phase that lasts approximately one diffusion time
$\tau\sim R^2/\eta^{\rm{eff}}$, the system settles in an eigenstate
characterized by exponential growth, respectively decay, 
$|\vec{B}|\sim e^{\gamma t}$ where $\gamma$ 
denotes the field amplitude growth rate.
\begin{figure}[h!] 
\includegraphics[width=0.47\textwidth]{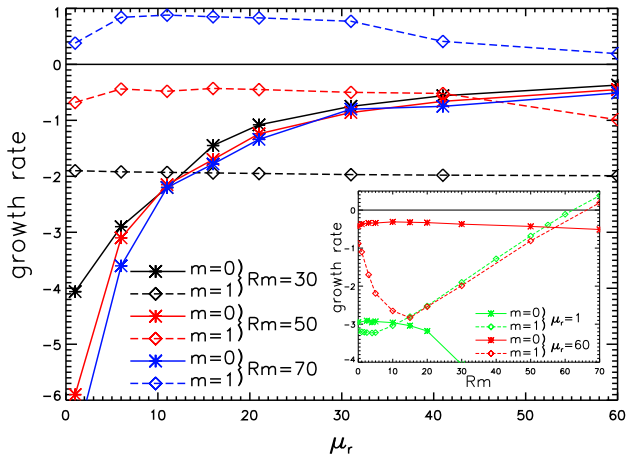}
\caption
{Growth rates of the field amplitude (without $\alpha$-effect)
Solid curves represent the results for the axisymmetric mode ($m=0$) and dashed
  curves show the $(m=1)$-mode. The inset plot presents the growth rates
  for $\mu_{\rm{r}}=1$ and $\mu_{\rm{r}}=60$ in dependence on ${\rm{Rm}}$.}
\label{fig::gr}
\end{figure}
Fig.~\ref{fig::gr} shows the growth rates for the axisymmetric
field mode ($m\!=\!0$, solid curves) and the simplest non-axisymmetric mode
$B\!\propto\!\!\cos\varphi$ ($m\!=\!1$, dashed curves).
Even at moderate values of $\mu_{\rm{r}}$ the $(m=0)$-growth rate is considerably
shifted towards the dynamo threshold (but remains still negative). The
reduction of this decay rate is roughly in agreement with
the reduction of $\eta_{\rm{eff}}$.
For large permeabilities the growth rate of the 
axisymmetric mode
saturates slightly below the dynamo threshold.
In the experimental
relevant interval (${\rm{Rm}}\!\la\!50$) the ($m\!=\!0$) growth rate dominates and
remains nearly independent of the applied ${\rm{Rm}}$. 
Contrary to the axisymmetric mode, the ($m=1$)-mode is hardly affected
by the permeability distribution (for ${\rm{Rm}}\ga 15$), and
for sufficiently large ${\rm{Rm}}$, the
($m=1$)-mode dominates again, resulting in dynamo action around
${\rm{Rm}}^{\rm{c}}\approx 60$.  
Preliminary simulations applying only an axisymmetric distribution for the
permeability (modeling the impeller disc without blades) exhibit a reduction of
the $(m=1)$-mode to ${\rm{Rm}}^{\rm{c}}\approx 50$ whereas the axisymmetric
field mode is hardly influenced (for ${\rm{Rm}}\la 50$).

It is important to note that so far no growing axisymmetric solutions were
obtained. 
One might have expected 
a sort of Busse-Wicht dynamo 
with an axisymmetric component modulated by a (unobserved) 
higher azimuthal
wavenumber
due to the azimuthally varying blade system.
But evidently, this mechanism alone
is not sufficient to explain the occurrence of the dominant axisymmetric
contribution in the VKS experiment.
In the following we will see that 
an axisymmetric growing solution can be obtained 
if an additional small $\alpha$-effect is invoked.
\begin{figure}[h!] 
\includegraphics[width=0.47\textwidth]{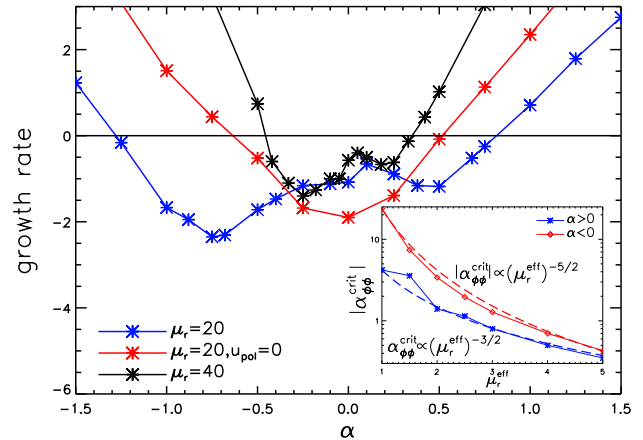}
\caption
{Growth rates of the axisymmetric mode for ${\rm{Rm}}=30$ in dependence of a
  prescribed homogenous $\alpha$-effect ($\alpha_{\varphi\varphi}$). 
  In all cases the (m=1)-mode lies below the growth rates of the
  axisymmetric solution.
  Additionally, the hypothetical case with
  $\vec{u}_{\rm{pol}}\!\!=\!\!u_r\vec{e}_r\!+\!u_z\vec{e}_z\!\!=\!\vec{0}$ has 
  been studied for $\mu_{\rm{r}}=20$.
  The inset plot shows the scaling behavior of the critical value for
  $\alpha_{\varphi\varphi}$ in dependence on $\mu_{\rm{r}}^{\rm{eff}}$. 
} 
\label{fig::gr_vs_alpha}  
\end{figure}
For the sake of simplicity we will assume 
a uniform distribution of 
$\alpha_{\varphi\varphi}$. 
This component is responsible for the generation of the 
poloidal field from
the toroidal field whereas the generation of the toroidal 
field essentially is determined
by shear
so that the other 
components, $\alpha_{rr}$ and
$\alpha_{zz}$, are assumed to be less important.
Because the high permeability impellers have already
shifted the growth rate of the ($m=0$)-mode 
very close to zero, it is not surprising that
a small $\alpha$-effect is now  sufficient to excite a 
growing axisymmetric field even at moderate values of the impeller
permeability (see Fig.~\ref{fig::gr_vs_alpha}).   
For larger $\mu_{\rm{r}}$ the dependence of the growth rates on the 
parameter $\alpha$ is rather delicate and even a small variation 
in $\alpha$ results in large changes of the corresponding growth rates. 
Growing axisymmetric fields are generated for both signs of $\alpha$, although
they are obtained a bit more easily for positive values of $\alpha$. 
The critical value of $\alpha_{\varphi\varphi}$ that is necessary for growing axisymmetric fields is
strongly decreasing for increasing $\mu_{\rm{r}}$
(see inset plot in Fig.~\ref{fig::gr_vs_alpha}).
In physical units the critical value for $\mu_{\rm{r}}=40$ corresponds to
$|\alpha^{\rm{c}}|\approx 0.15 
\mbox{ ms}^{-1}$ which is much lower than the critical value reported in
\cite{andregafd} where $|\alpha^{\rm{crit}}|\approx 16 \mbox{
ms}^{-1}$ was obtained (for $\mu_{\rm{r}}=1$ and ${\rm{Rm}}= 32$). 
The local maxima for $\mu_{\rm{r}}=20$ (blue curve) and
${\mu_{\rm{r}}}=40$ (black curve) close to $\alpha=0$ correspond to
stationary solutions whereas a transition to oscillating
solutions occurs approaching the adjacent minima which might be the cause for the
local suppression of the growth rate.
Interestingly, the hypothetical omission of the poloidal velocity component
($\vec{u}_{\rm{pol}}\!\!=\!u_r\vec{e}_r\!\!+\!u_z\vec{e}_z\!\!=\!\!{\vec{0}}$) 
works even in favor of the dynamo (see red curve in Fig.~\ref{fig::gr_vs_alpha}). 
This might explain the intriguing experimental observation that, despite a
careful optimization of the impeller blades for one rotation direction, a
change of this blade orientation does not alter significantly the critical
${\rm{Rm}}$ \cite{2009PhFl...21c5108M}. 
As long as the $\alpha$-effect close to the impellers is strong enough,
poloidal field generation is sufficient to sustain the dynamo cycle. 
The resulting magnetic field is strongly concentrated in the impeller
region (see Fig.~\ref{fig::isoener}a).
For larger ${\rm{Rm}}$, the field geometry is determined by the well known
($m=1$) dominated field structure (Fig. 4b) and dynamo action
predominantly takes place in the bulk of the cylinder which explains the
impassivity of the ($m=1$)-mode towards the high $\mu_{\rm{r}}$ disks located
close to the end caps.   
\begin{figure}[h!] 
\includegraphics[width=0.49\textwidth]{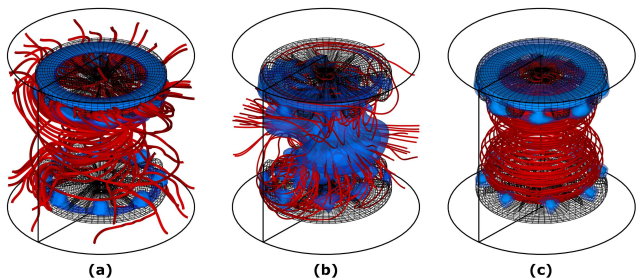}
\caption
{Distribution of magnetic energy density (blue colored isosurfaces at 10\% of
  the maximum value) and geometric structure of the magnetic
field (red streamlines) for
  $\mu_{\rm{r}}=20$: 
(a) ${\rm{Rm}}=30, \alpha=-1.5$, growing axisymmetric mode;
(b) ${\rm{Rm}}=70, \alpha=0$, growing $(m=1)$-mode; 
(c) ${\rm{Rm}}=30, \alpha=0$, decaying axisymmetric mode. 
Note the ($m=8$) modulation induced by the impeller blades.}
\label{fig::isoener}
\end{figure}
Note that in case of a vanishing $\alpha$-effect the (decaying) solution is dominated
by an axisymmetric toroidal field and the poloidal field generation is not
sufficient to close the dynamo cycle (Fig.~\ref{fig::isoener}c). 
At this point, a deeper investigation requires a more 
detailed modeling of the $\alpha$-effect on the basis of water experiments.
Direct numerical simulations of forced turbulence are far away from
realistic parameters so that reliable models for the $\alpha$-distribution
can hardly be extracted from DNS models.
Preliminary simulations with a different spatial distribution of
$\alpha$ suggest that the extreme facilitation of the axisymmetric 
mode by the soft-iron disks is a generic effect which takes place as long as
sufficient poloidal field generation (due to $\alpha$) occurs nearby the
impellers.

The dynamo process is strongly influenced by the reduction of $\eta^{\rm{eff}}$. 
However, the decrease of $\eta^{\rm{eff}}$ is not sufficient to explain the
results because simulations with the permeability
variation replaced by a conductivity variation of the same magnitude (and
space/time dependence) exhibit a significant deviation
in the magnitude of the growth rates (although the tendency remains similar).

Our results clearly demonstrate that the common simplified treatment 
of the high permeability impellers, by demanding vanishing tangential
field components at the interface, is not justified.
In the relevant range of ${\rm{Rm}}$, the high 
permeability works selectively in favor 
of the ($m=0$)-mode while the ($m=1$)-mode is 
only slightly affected.
To initiate a growing $(m=0)$-mode,
some $\alpha$-effect
has still to be invoked, although the scaling behavior with $\mu_{r}$ 
indicates that its magnitude is very small. 
The homogenous $\alpha$-distribution should be regarded as the simplest
example how the soft iron disks facilitate growing axisymmetric solutions at  
reasonable parameter values.
A complementary and more detailed approach will have to consider a
non-axisymmetric flow variation 
in terms of azimuthally drifting equatorial vortices  
that have been observed in water experiments
reported by \cite{2007PhRvL..99e4101D}. 
Similar as in the $\alpha$-model with high permeability blades this
should lead to a mixed mode solution of the magnetic field where a higher
azimuthal mode is present.  
\acknowledgments{We acknowledge fruitful discussions with A. Chiffaudel,
  F. Daviaud, S. Fauve, C. Gissinger, J. L\'eorat, C. Nore and
  J.-F. Pinton. 
Financial support from Deutsche Forschungsgemeinschaft
  in frame of the Collaborative Research Center SFB609 is gratefully
  acknowledged.}

%-------------------- End of Body
\bibliographystyle{apsrev}

\end{document}